\documentclass[iop]{emulateapj}

\newcommand{\kms}{km s$^{-1}$}
\newcommand{\msun}{$M_{\sun}$}
\newcommand{\mbh}{$M_{BH}$}
\newcommand{\lsc}{$L_{sc}$}
\newcommand{\lin}{$L_{in}$}
\newcommand{\hubu}{km s$^{-1}$ Mpc$^{-1}$}
\newcommand{\flxu}{ergs cm$^{-2}$ s$^{-1}$ \AA$^{-1}$}
\newcommand{\lumu}{ergs s$^{-1}$}
\newcommand{\etal}{{\rm et al.\/\ }}

\newcommand{\firr}{$f_{25}/f_{60}$}
\newcommand{\ftv}{$f_{25}$}
\newcommand{\fst}{$f_{60}$}

\newcommand{\ha}{H$\alpha$}
\newcommand{\hb}{H$\beta$}

\newcommand{\te}{$T_e$}
\newcommand{\nee}{$n_e$}
\newcommand{\cc}{cm$^{-3}$}
\newcommand{\oiii}{{\rm [O~III]}}
\def\wave#1{$\lambda${#1}}

\slugcomment{Received 2009 October 15; accepted 2009 December 08}
\journalinfo{Accepted for publication in the Astrophysical Journal}

\shortauthors{TRAN}
\shorttitle{HIDDEN DOUBLE-PEAKED EMITTERS}

\begin{document}

\title{Hidden Double-Peaked Emitters in Seyfert 2 Galaxies}

\author{Hien D. Tran}
\affil{W. M. Keck Observatory, 65-1120 Mamalahoa Highway, Kamuela, HI 96743, USA}
\email{htran@keck.hawaii.edu}

\begin{abstract}
We present the detection of extremely broad, double-peaked, highly polarized
\ha~emission lines in the nuclei of the well-known Seyfert 2 galaxies NGC 2110 
and NGC 5252. These hidden broad \ha~emission lines, visible only in scattered light, 
are shown to display significant variability in strength and profile on timescales 
of $\lesssim$ 1 yr. That the broad emission line exhibits variability in polarized flux 
also suggests that the scattering region must be very compact, possibly confined in a small 
number of electron clouds $\lesssim$ 1 lt-yr in size. Our observational constraints place these 
clouds within $\sim$~10 pc of the nucleus with temperatures \te~$\lesssim$~$10^6$ K and densities 
\nee~$\sim 10^{7}$ \cc, consistent with a region just outside the obscuring torus 
between the broad-line region and narrow-line region. These scattering clouds could arise from 
the clumpy torus itself. 
These findings and other properties indicate that NGC 2110 and NGC 5252 are the hidden 
counterparts to the broad-line 
double-peaked emission-line active galactic nuclei, whose examples include Arp 102B and 3C 332. 
\end{abstract}


\section{Introduction} \label{intro}

\object{NGC 5252} ($z$ = 0.023) and \object{NGC 2110} ($z$ = 0.0078) are two nearby, well-known and 
well-studied Seyfert 2 galaxies (S2s). Both are known to 
have very extended, well-defined, and spectacular ionization cones (e.g., Mulchaey et al. 1994; 
Wilson \& Tsvetanov 1994). Along with the high-ionization 
lines typical of Seyfert 2s, their spectra also show unusually strong low-ionization lines more 
typical of LINERs \citep{hs83,gon98}. NGC 2110 was first recognized as a very strong X-ray source 
\citep{br78} and has been studied extensively at X-ray and other wavelengths 
(e.g., Evans et al. 2006). Both galaxies have been popular and frequent targets as tests of 
the active galactic nucleus (AGN) unification model \citep{a93}, and there have been previous 
searches for hidden broad-line regions (HBLRs) in both of these objects. 
However, the results were either negative \citep{kay94} or 
detection of broad \ha~only ``marginal'' with large uncertainty \citep{y96}.
It was not until recently that NGC 2110 was found to show highly polarized double-peaked 
\ha~emission by \citet{m07}. In this paper, we confirm the double-peak nature of the hidden broad 
\ha~in NGC 2110 and also report the discovery of a similar double-peaked emission line profile in
the polarized broad \ha~from the nucleus of NGC 5252. 
As part of a survey of Seyfert galaxies, both objects have been monitored with 
the polarimeter on the low-resolution spectrometer at the W. M. Keck Observatory (WMKO). 
We report here the first results of the spectropolarimetric monitoring of these objects.
Throughout this paper, we assume $H_o$ = 75 \hubu, $q_o$ = 0 and $\Lambda$ = 0. At the distances of 
NGC 2110 (31.3 Mpc) and NGC 5252 (93.0 Mpc), 1\arcsec~corresponds to a projected size of 150 pc and 
430 pc, respectively.

\section{Observations} \label{obs}

Spectropolarimetric observations were made with the low resolution imaging spectrograph 
(LRIS; \citet{oke95}) and polarimeter on the 10-m Keck I telescope. We used a 1\arcsec~long 
slit centered on the nucleus of the AGN.  The slit was oriented along the cone axes at 
position angle, P.A. = 160\arcdeg~in NGC 2110 and P.A. = 165\arcdeg~in NGC 5252 to include the extended 
emission-line structures. 
We used a 300 grooves mm$^{-1}$ grating with the red arm of LRIS, giving a dispersion of 
2.46~\AA~pixel$^{-1}$ and resolution of $\sim$ 10 \AA~(FWHM), covering a wavelength range of 
$\sim$ 3900--8900 \AA. 
The observations were made by following standard procedures of rotating
the half waveplate to four position angles (0\arcdeg, 22\fdg5, 45\arcdeg,
and 67\fdg5), and dividing the exposure times equally among them. 
We obtained five epochs of observations for each of the two targets, covering a period of approximately four
years for NGC 5252 and approximately two years for NGC 2110, as shown in the log of observations in Table 
\ref{olog}.  Flux, polarization and null standard stars were observed each night for data 
calibration.

\begin{deluxetable*}{lccccccc}[t]
\tablecolumns{8}
\tabletypesize{\scriptsize}
\tablewidth{0pt}
\tablecaption{Journal of Observations \label{olog}}
\tablehead{
\colhead{Object} & \colhead{UT Date} & \colhead{Exposures} & \colhead{$P_c^a$} & \colhead{$\theta_c^a$} & \colhead{$P_{H\alpha}^b$} & \colhead{$\theta_{H\alpha}^b$} & \colhead{Epoch} \\
\colhead{} & \colhead{} & \colhead{(s)} & \colhead{(\%)} & \colhead{(\arcdeg)} & \colhead{(\%)} & \colhead{(\arcdeg)} & \colhead{} 
}
\startdata
NGC 2110 & 2006 Jan 25 & 4$\times$\ 900& 0.18 $\pm$ 0.03 & 67.9 $\pm$ 4.0 & 3.5 $\pm$ 0.09 & 67.4 $\pm$ 0.7 & 1 \\
         & 2006 Dec 17 & 4$\times$1000& 0.48  $\pm$ 0.02 & 67.5 $\pm$ 1.2 & 2.2 $\pm$ 0.09 & 68.8 $\pm$ 0.9 & 2 \\
         & 2007 Feb 15 & 4$\times$1500& 0.45  $\pm$ 0.01 & 74.8 $\pm$ 0.8 & 1.9 $\pm$ 0.07 & 69.9 $\pm$ 0.9 & 3 \\
         & 2007 Nov 17 & 4$\times$1200& 0.54  $\pm$ 0.01 & 70.0 $\pm$ 0.7 & 2.9 $\pm$ 0.11 & 67.7 $\pm$ 1.0 & 4 \\
         & 2008 Mar 12 & 4$\times$1500& 0.31  $\pm$ 0.02 & 84.4 $\pm$ 1.8 & 2.2 $\pm$ 0.07 & 69.3 $\pm$ 1.0 & 5 \\
NGC 5252 & 2004 Jun 17 & 4$\times$1000& 1.60  $\pm$ 0.02 & 80.1 $\pm$ 0.3 & 2.9 $\pm$ 0.10 & 72.5 $\pm$ 0.9 & 1 \\
         & 2005 May 14 & 4$\times$1200& 1.40  $\pm$ 0.02 & 82.6 $\pm$ 0.4 & 2.1 $\pm$ 0.09 & 75.3 $\pm$ 1.1 & 2 \\
         & 2007 Feb 15 & 4$\times$1350& 1.58  $\pm$ 0.02 & 78.7 $\pm$ 0.3 & 2.8 $\pm$ 0.08 & 72.6 $\pm$ 0.8 & 3 \\
         & 2007 Apr 12 & 4$\times$1200& 1.69  $\pm$ 0.02 & 79.3 $\pm$ 0.3 & 3.1 $\pm$ 0.08 & 72.9 $\pm$ 0.7 & 4 \\
         & 2008 Mar 12 & 4$\times$1200& 1.92  $\pm$ 0.02 & 75.1 $\pm$ 0.3 & 4.5 $\pm$ 0.07 & 70.0 $\pm$ 0.5 & 5 \\
\enddata
\tablenotetext{a}{Observed average continuum polarizations ($P$) and position angles ($\theta$) over the observed wavelength ranges 5100--6300\AA~for NGC 2110, and 5200--6500\AA~for NGC 5252, which are relatively free of strong emission lines.}
\tablenotetext{b}{Observed average polarizations in the wings of broad \ha~emission line. For NGC 2110,
the average is in the {\it blue} wing over the observed wavelength range 6457-6501\AA. For NGC 5252, the 
average is in the {\it red} wing over the observed wavelength range 6761-6812\AA.} 
\end{deluxetable*}

Spectropolarimetric data reduction was done with 
standard techniques using a combination of IRAF and VISTA, as described in, e.g., Tran (1995a). 
The data were extracted using apertures 11 and 13 pixels wide, corresponding to 2\farcs3 and 2\farcs7 
surrounding the nucleus of NGC 5252 and NGC 2110, respectively. 

In order to get a good handle on the interstellar polarization (ISpol) in our Galaxy, following the 
prescription of \citet{t95}, several stellar probes near the line of sight to these objects were 
selected and observed with the same instrumental setups. The results for these probes are shown in 
Table \ref{isp}.

\begin{deluxetable*}{lccccccr}
\tablecolumns{8}
\tabletypesize{\scriptsize}
\tablewidth{0pt}
\tablecaption{Probes of Interstellar Polarizations \label{isp}}
\tablehead{
\colhead{Object} & \colhead{$b_{II}^a$} & \colhead{$E(B-V)^b$} & \colhead{Probes$^c$} & \colhead{Separation$^d$} & \colhead{Distance$^e$} & \colhead{$P^f$} & \colhead{$\theta^f$} \\
\colhead{} & \colhead{(\arcdeg)} & \colhead{$P_{max}$} & \colhead{} & \colhead{(\arcmin)} & \colhead{(pc)} & \colhead{(\%)} & \colhead{(\arcdeg)} 
}
\startdata
NGC 2110  & $-$16.5  & 0.375  & PPM 188547 (1) & 5.4  & 270 & 0.44  & 55.6 \\
          & (528)    & 3.37\% & PPM 702444 (2) & 13.0 & 524 & 0.27  & 22.6 \\
          &          &        & PPM 188546 (3) & 36.5 & 380 & 0.36  & 65.6 \\
          &          &        & PPM 188568 (4) & 50.5 & 549 & 0.11  & 165. \\
NGC 5252  & 64.8     & 0.034  & PPM 159913 (1) & 16.7 & 180 & 0.12  & 71.6 \\
          & (165)    & 0.31\% & PPM 159939 (2) & 35.1 & 100 & 0.21  & 73.4 \\
          &          &        & PPM 159899 (3) & 40.7 & 260 & 0.16  & 73.8 \\
          &          &        & PPM 159919 (4) & 46.9 & 200 & 0.14  & 80.5 \\
\enddata
\tablenotetext{a}{Galactic latitude. Value in parentheses denotes the minimum distance in pc at which 
the probe should lie; $d = 150\ {\rm csc}\ b_{II}$.}
\tablenotetext{b}{Galactic interstellar reddening from Schlegel et al. (1998). The maximum expected ISpol,
 $P_{max}$ obeys the relation $P_{max} \leq 9E(B-V)$ \citep{ser75}.}
\tablenotetext{c}{Number in parentheses denotes star number plotted in Figure \ref{contpol}.}
\tablenotetext{d}{Spatial separation between the probe and object in the plane of the sky.}
\tablenotetext{e}{Approximate distance of star from spectroscopic parallax.}
\tablenotetext{f}{Observed average over the wavelength range 4600--6800\AA.}
\end{deluxetable*}

\section{Results} \label{res}

\subsection{Spectropolarimetry and Interstellar Polarization}

\begin{figure}[h]
\plotone{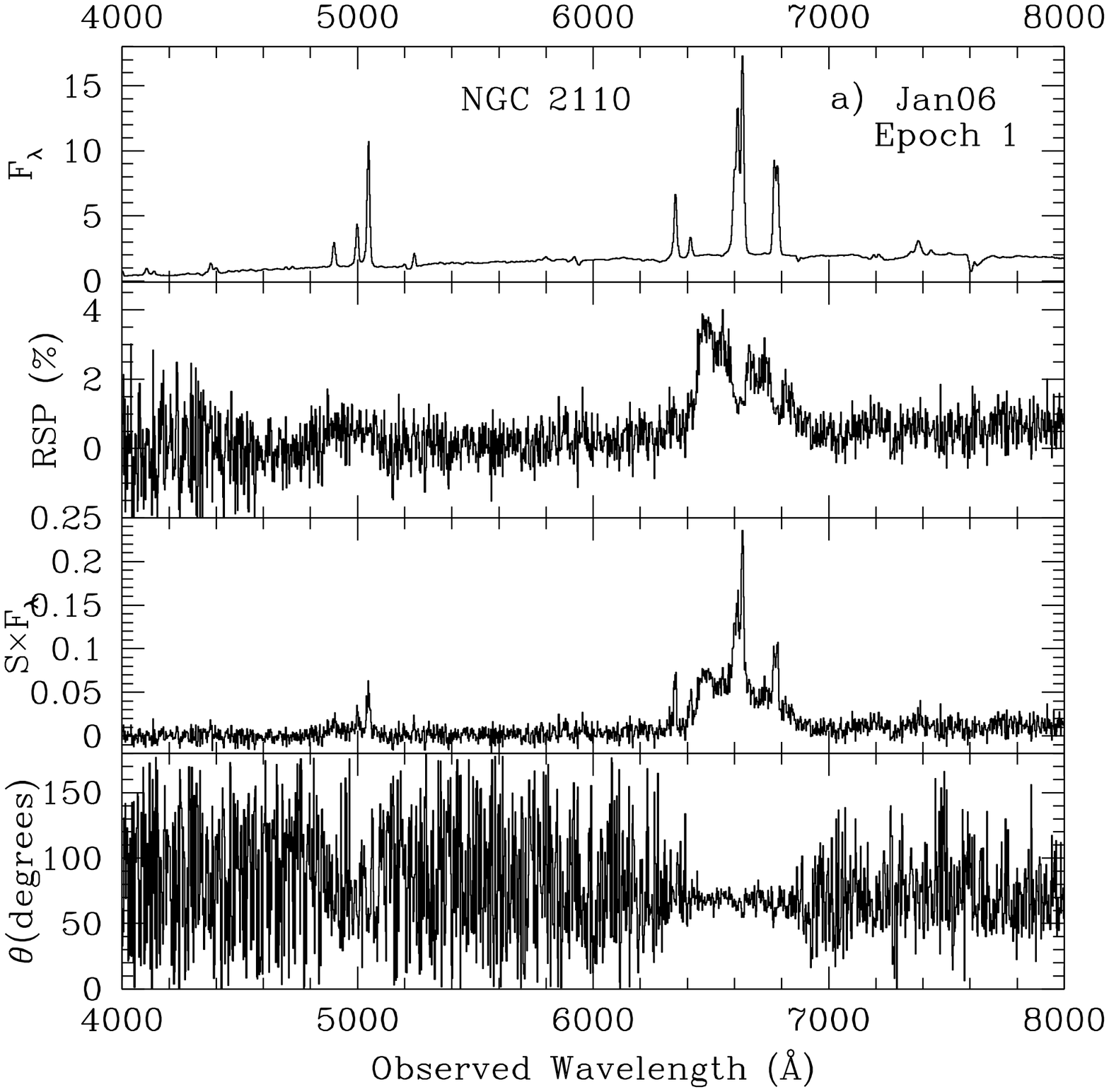}
\caption{Observed spectropolarimetry of NGC 2110 for all five epochs. {\it (a)} Epoch 1: 2006 January,
{\it (b)} epoch 2: 2006 December, {\it (c)} epoch 3: 2007 February, {\it (d)} epoch 4: 2007 November, and
{\it (e)} epoch 5: 2008 March. From top to bottom are the 
total flux spectrum, observed degree of polarization, presented as rotated Stokes parameter (RSP), 
polarized flux, or Stokes flux spectrum ($S\times F_\lambda$), 
and polarization P.A. ($\theta$). The flux scales are in units of 10$^{-15}$ \flxu.
\label{n2110}}
\end{figure}

\begin{figure}[h]
\plotone{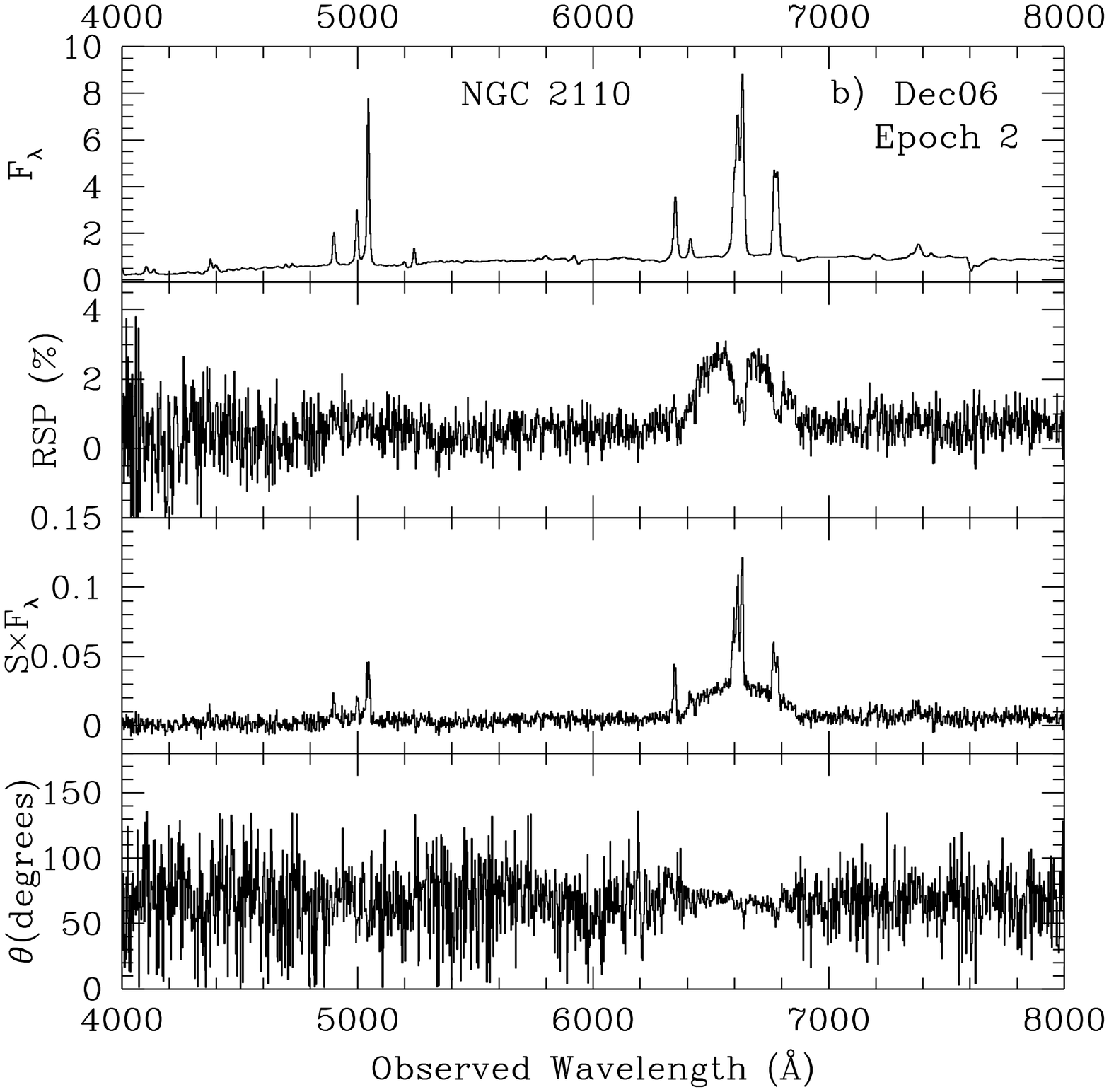}
\end{figure}

\begin{figure}[h]
\plotone{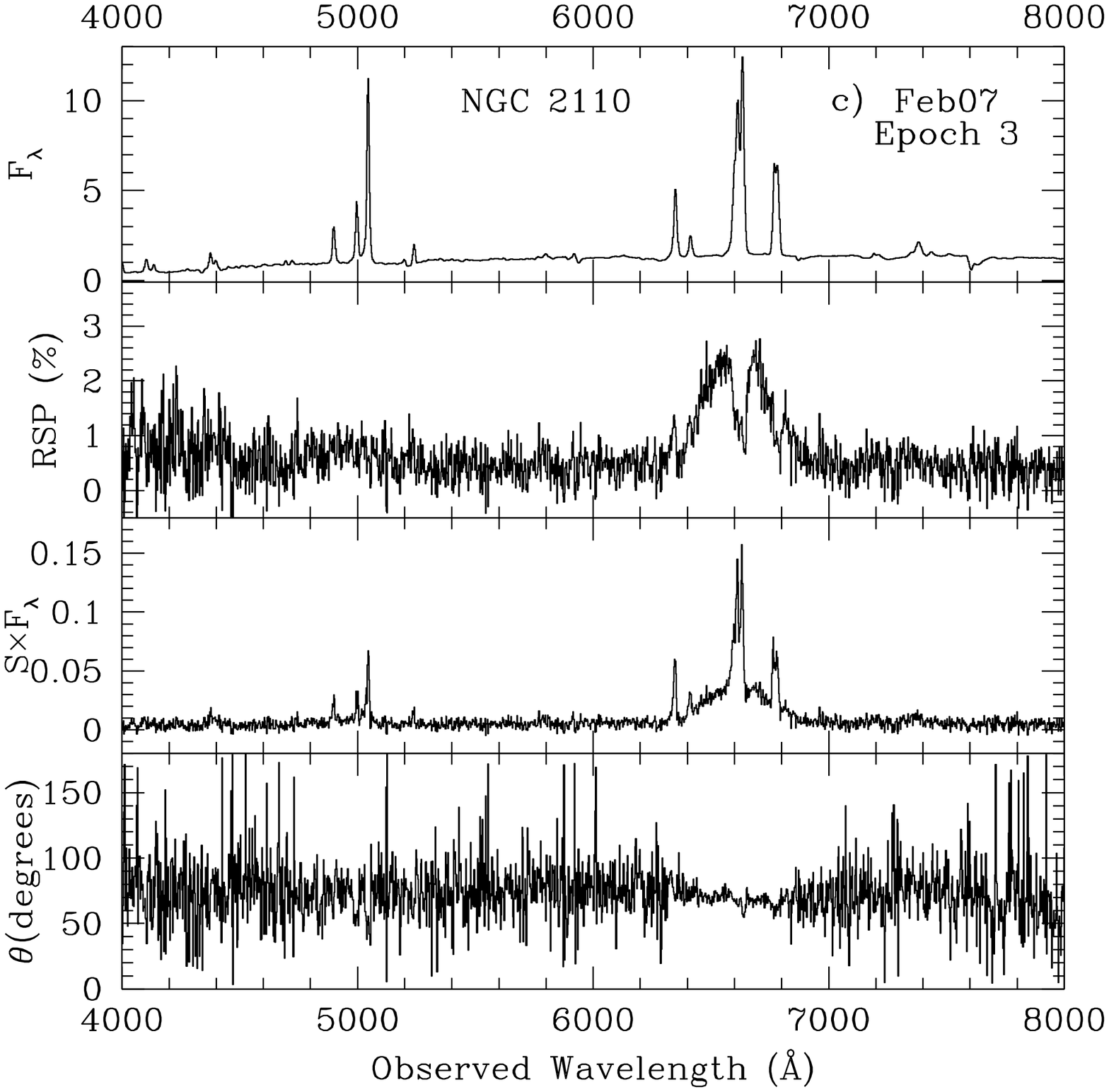}
\end{figure}

\begin{figure}[h]
\plotone{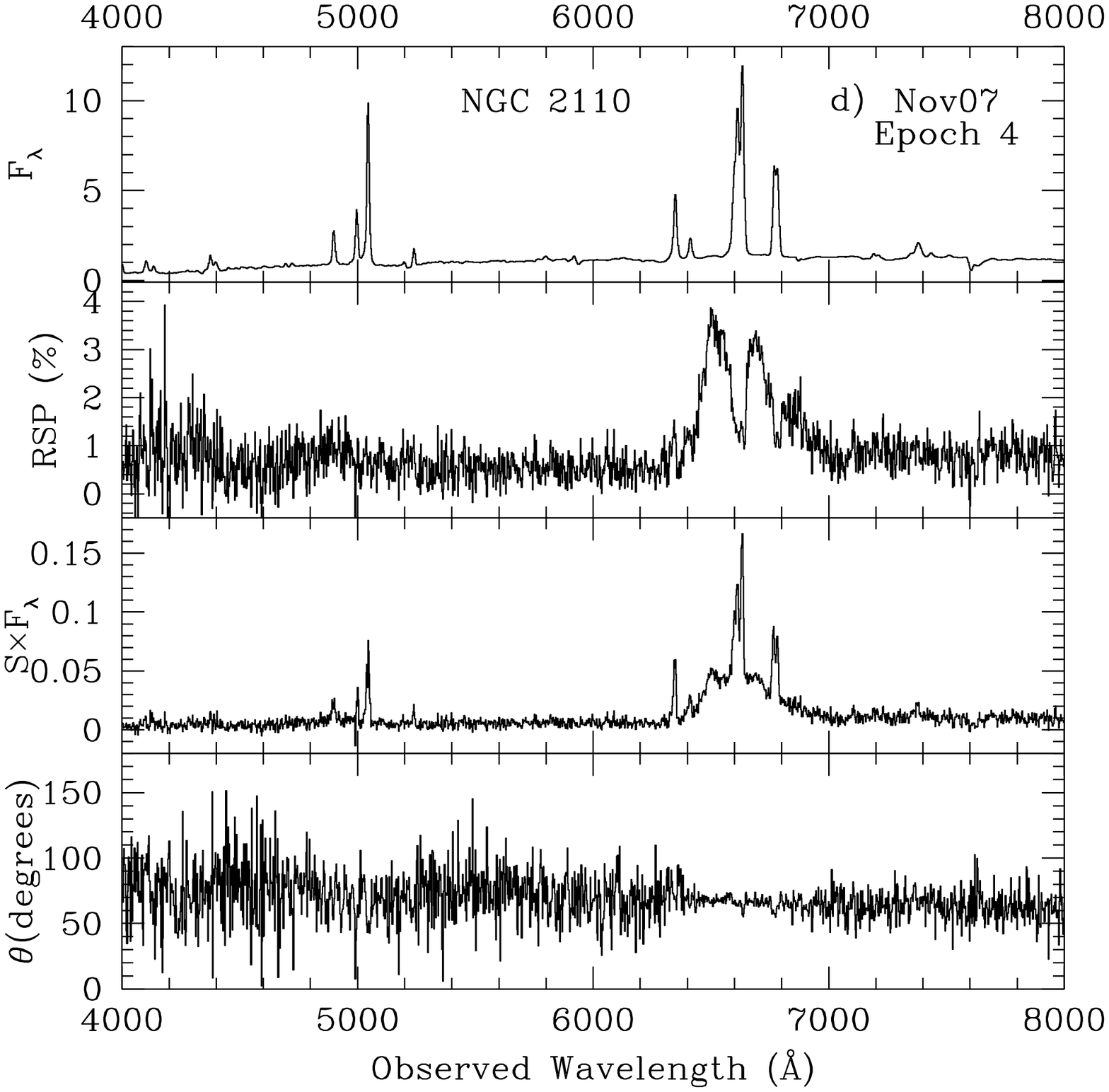}
\end{figure}

\begin{figure}[h]
\plotone{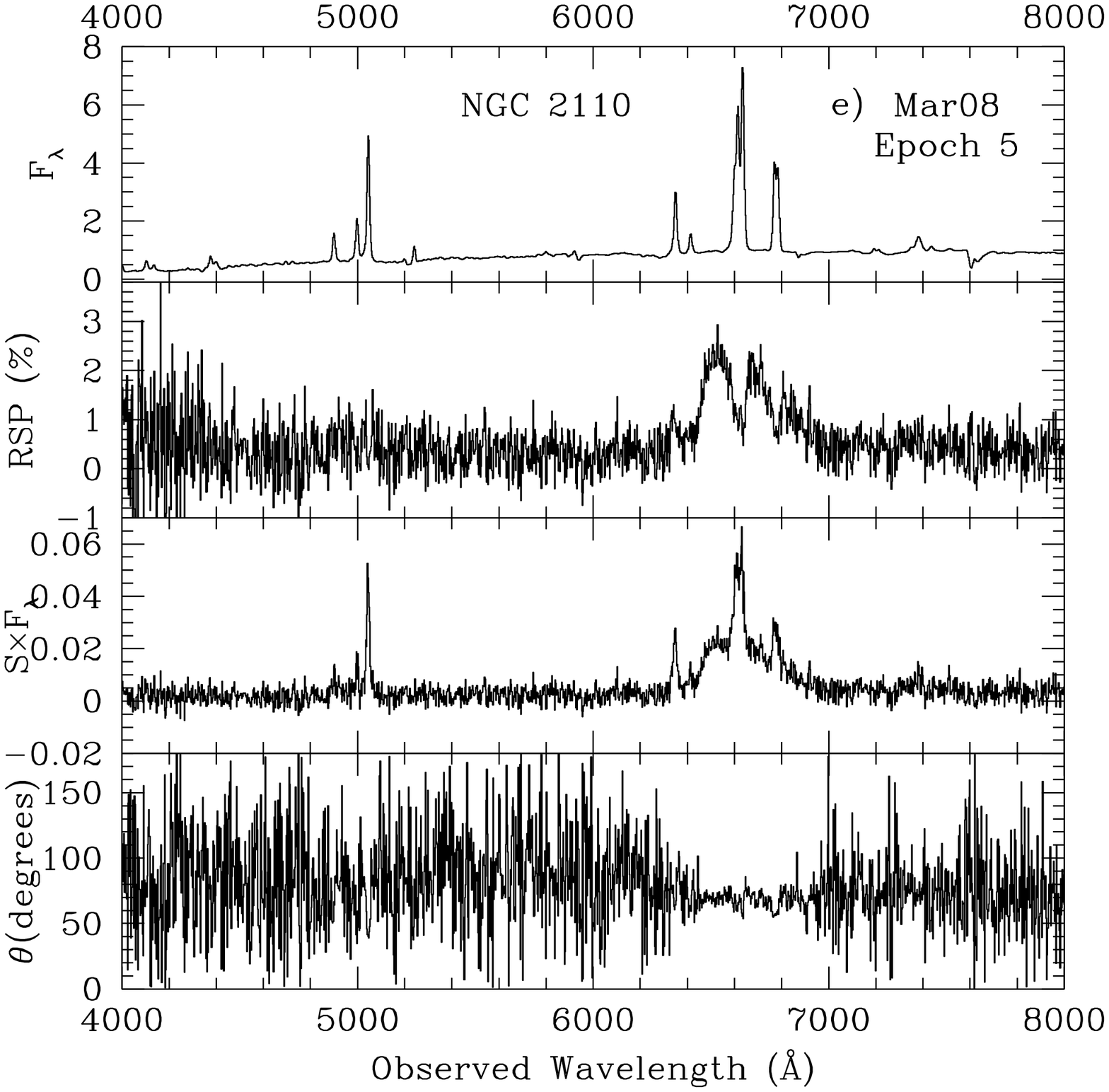}
\end{figure}

\begin{figure}
\plotone{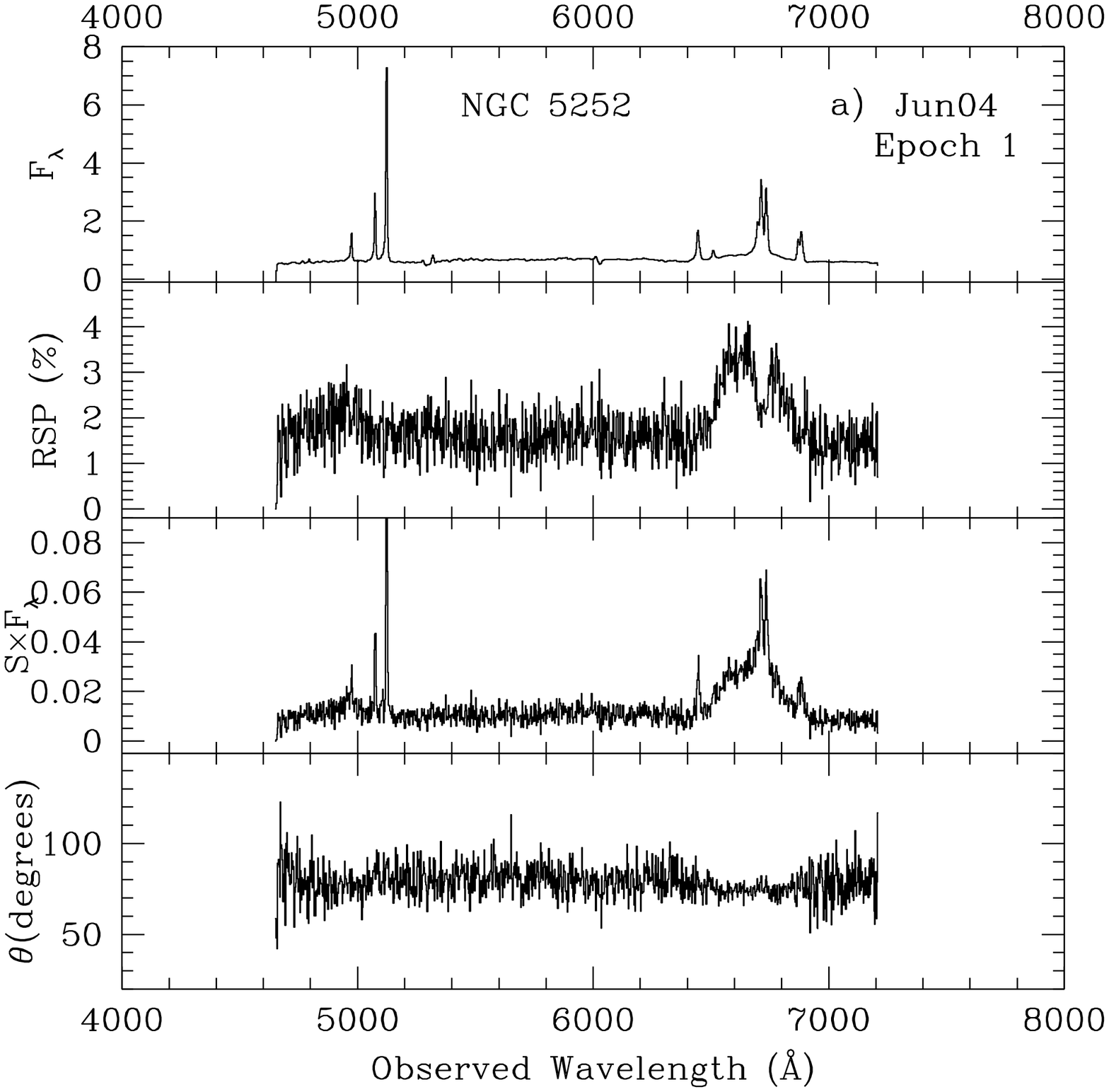}
\caption{Observed spectropolarimetry of NGC 5252 for all five epochs, arranged as in Figure \ref{n2110}.
{\it (a)} Epoch 1: 2004 June,
{\it (b)} epoch 2: 2005 May, {\it (c)} epoch 3: 2007 February, {\it (d)} epoch 4: 2007 April, and
{\it (e)} epoch 5: 2008 March.
\label{n5252}}
\end{figure}

The spectropolarimetry for NGC 2110 and NGC 5252 for all epochs is presented in Figures \ref{n2110} 
and \ref{n5252}, respectively. 
The observed polarization is 
very high in the broad emission lines, peaking at several percent, but drops in the narrow lines and is
quite low in the continuum, being only $\sim$0.5\% in NGC 2110. As can be seen, both objects display 
spectacularly broad \ha~emission lines in the polarized flux spectra. Although such polarized broad 
lines have been seen before in S2s, what is remarkable is that these hidden broad lines are among 
the broadest ever observed, with FWHM $\sim$ 13,000 - 17,000 \kms and FWZI $\sim$~25,000 - 32,000 \kms.
Note also that these broad line profiles are asymmetric and reminiscent of what is seen in the 
double-peaked emission line AGNs (see e.g., Eracleous 2004). Although broad \ha~is very prominent in
the polarized flux spectra of both galaxies, broad \hb~is weak or absent, a feature also noted 
by \citet{m07}. This may imply that the polarized flux is highly reddened. 
In an attempt to detect the polarized broad \hb, we co-added the five epochs of polarized flux spectra
of each galaxy and present the averaged Stokes flux in Figure \ref{avepf}. Broad \hb~can now be easily 
seen in both objects. We measure a broad-line Balmer decrement \ha/\hb~of $\sim$ 10 for NGC 2110 and 
3.9 for NGC 5252. 
Although the broad-line Balmer decrement in AGNs has generally been thought to be difficult or unsuitable to use as a 
reddening indicator due to the extreme conditions of the BLRs that make it susceptible to collisional excitation 
and other radiative transfer effects \citep[e.g.,][]{ree89}, \citet{dong08} found that statistically, the mean intrinsic 
broad-line \ha/\hb~in a large sample of Seyfert 1s and QSOs is actually about 3, only slightly steeper than 
Case B value, with very little dispersion. Furthermore, they found that this ratio is rather insensitive to different AGN 
properties, with a mean empirical value for a sub-sample of double-peaked emitters (DPEs) to be 3.27. Assuming that this 
is the intrinsic Balmer decrement for this type of objects, we infer a reddening of $E(B-V) = 1.1$ ($A_V = 3.5$) 
and $E(B-V) = 0.18$ ($A_V = 0.55$) for the broad-line regions (BLRs) of NGC 2110 and NGC 5252, respectively.

\begin{figure}
\plotone{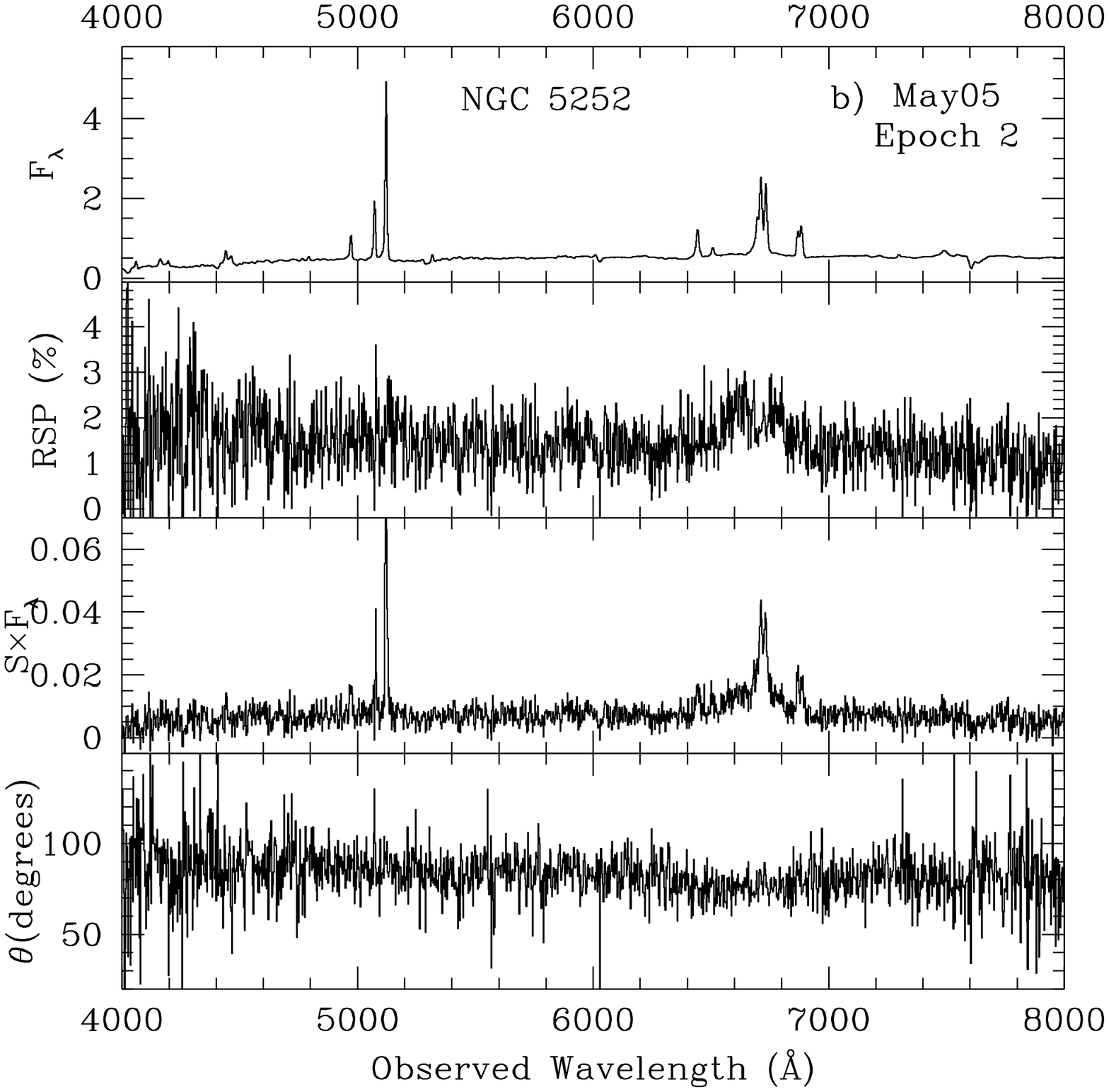}
\end{figure}

\begin{figure}
\plotone{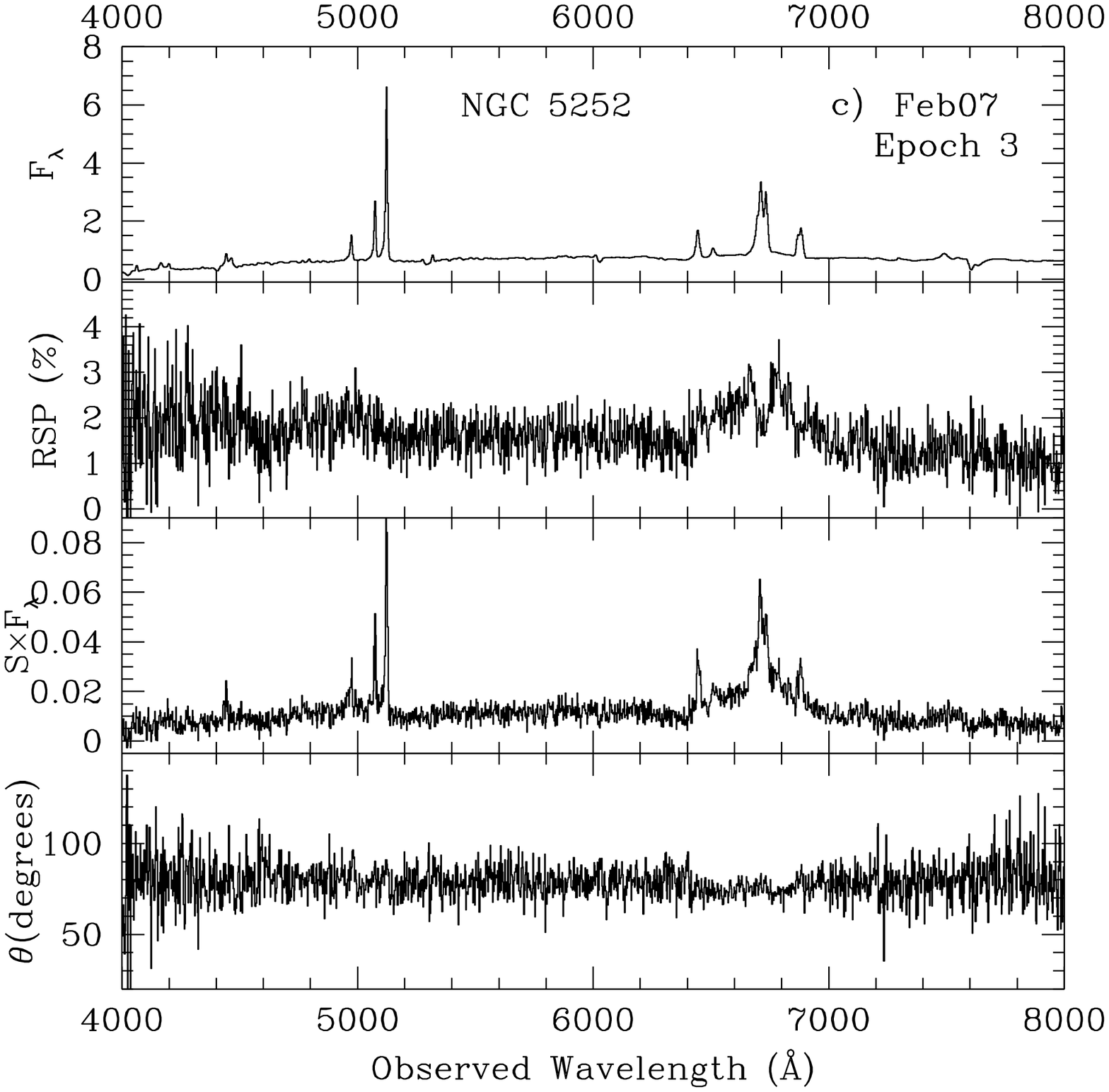}
\end{figure}

\begin{figure}
\plotone{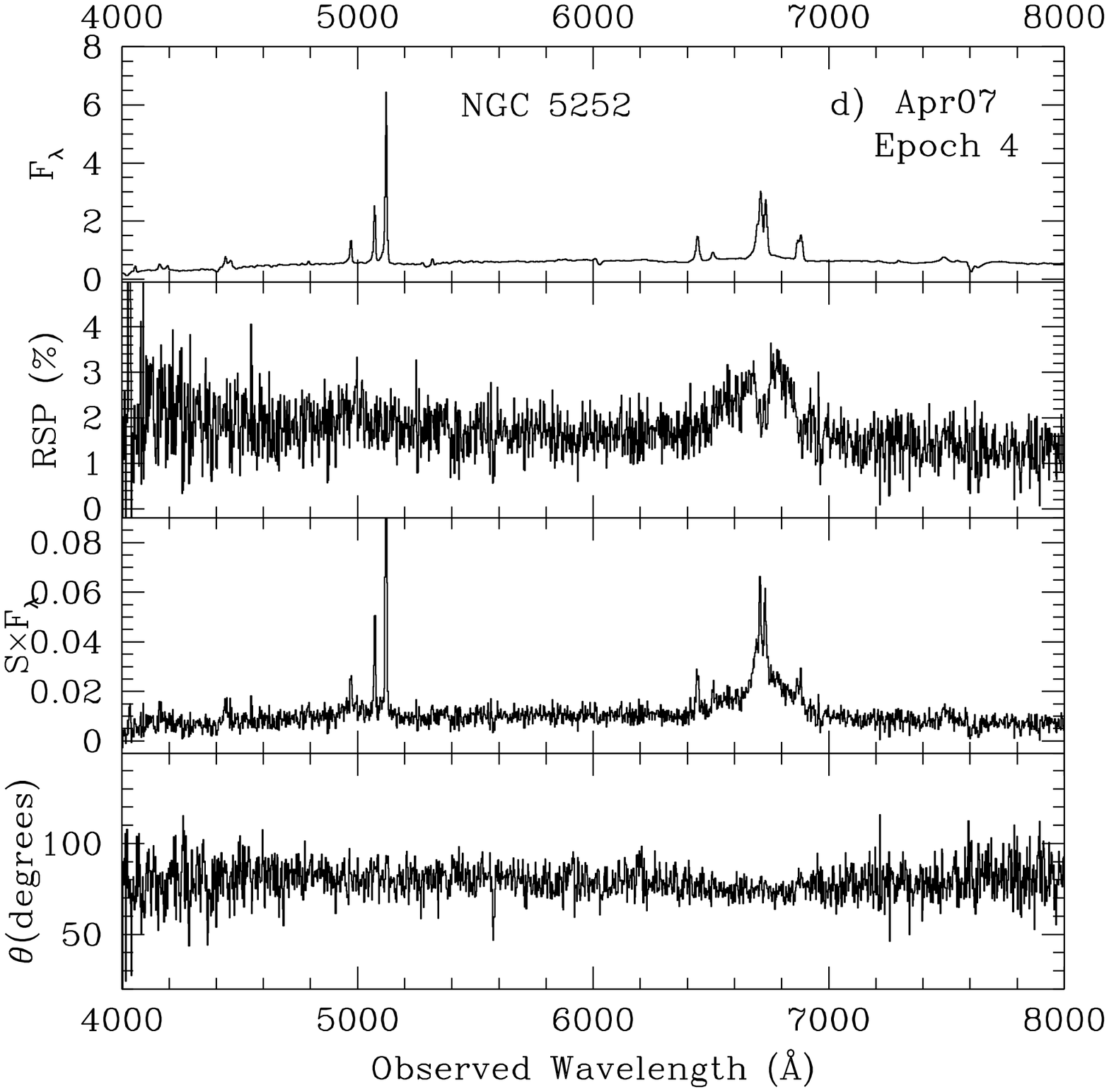}
\end{figure}

\begin{figure}
\plotone{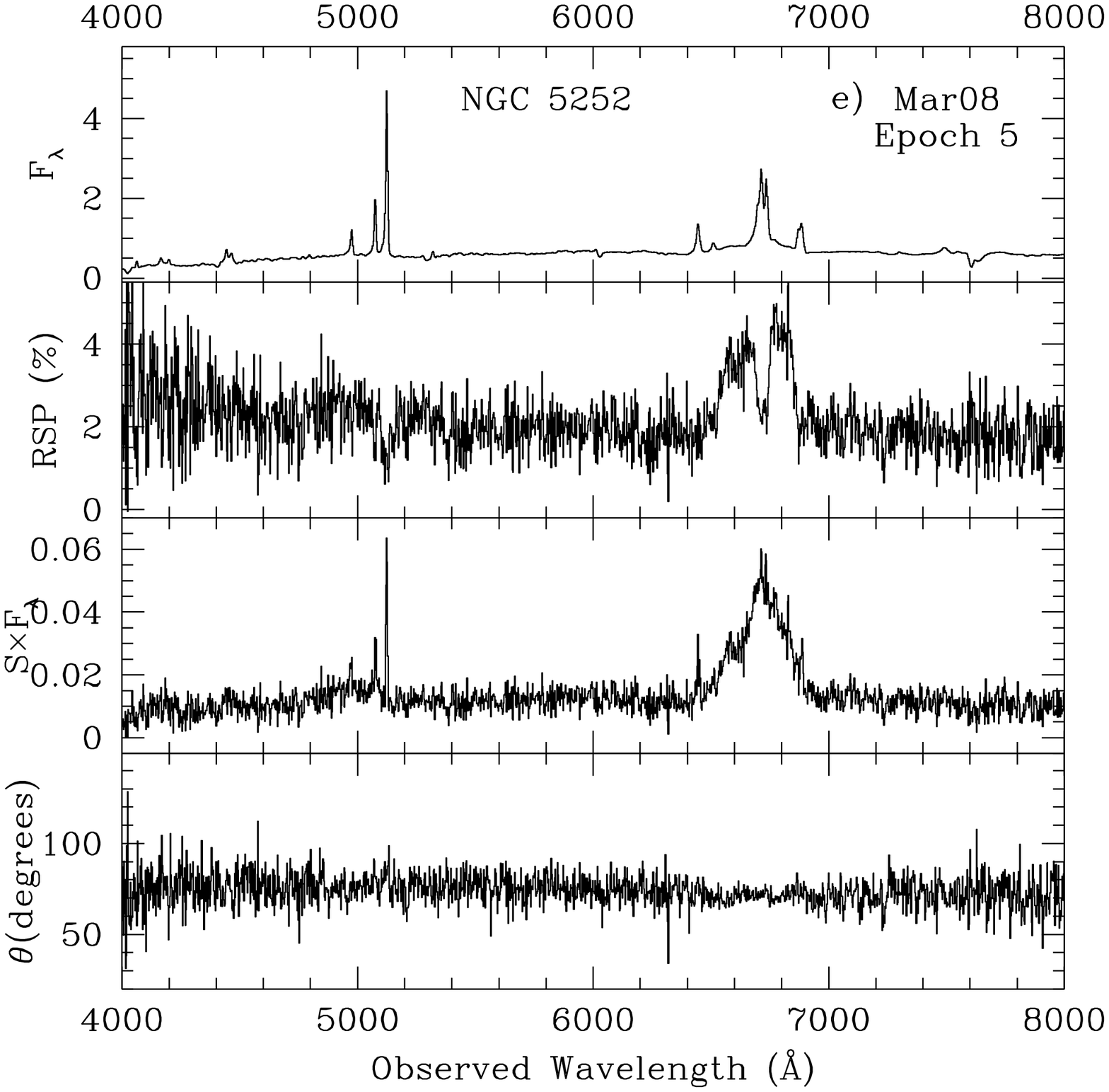}
\end{figure}

\begin{figure}
\plotone{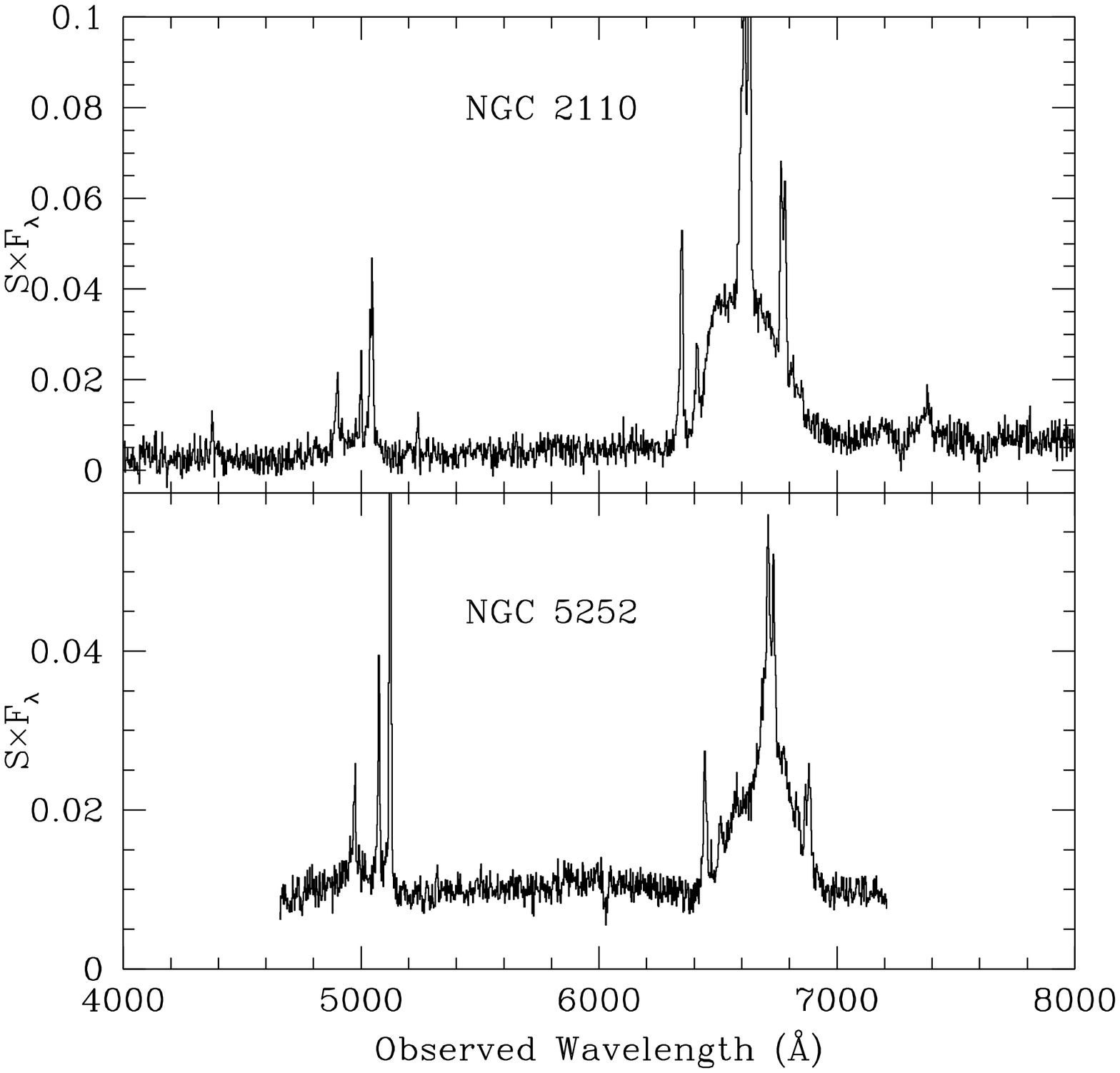}
\caption{Average Stokes flux spectra of NGC 2110 and NGC 5252 over all five epochs. The polarized broad 
\hb~emission line, hardly detectable in each single epoch observation, is clearly visible.   
\label{avepf}}
\end{figure}

Table \ref{olog} presents the mean observed polarization for the continuum between [O III] \wave 5007
and \ha, which is relatively free of strong emission lines for each epoch of observations. Also listed 
for comparison are the mean observed polarizations in the wings of the broad \ha~emission line.
We selected a region that displayed the highest peak observed polarization for each object. For
NGC 2110, it was the blue wing of \ha~in the observed wavelength range 6457-6501\AA, and for NGC 5252,
it was the red wing in the observed wavelength range 6761-6812 \AA. 

\begin{figure*}
\plotone{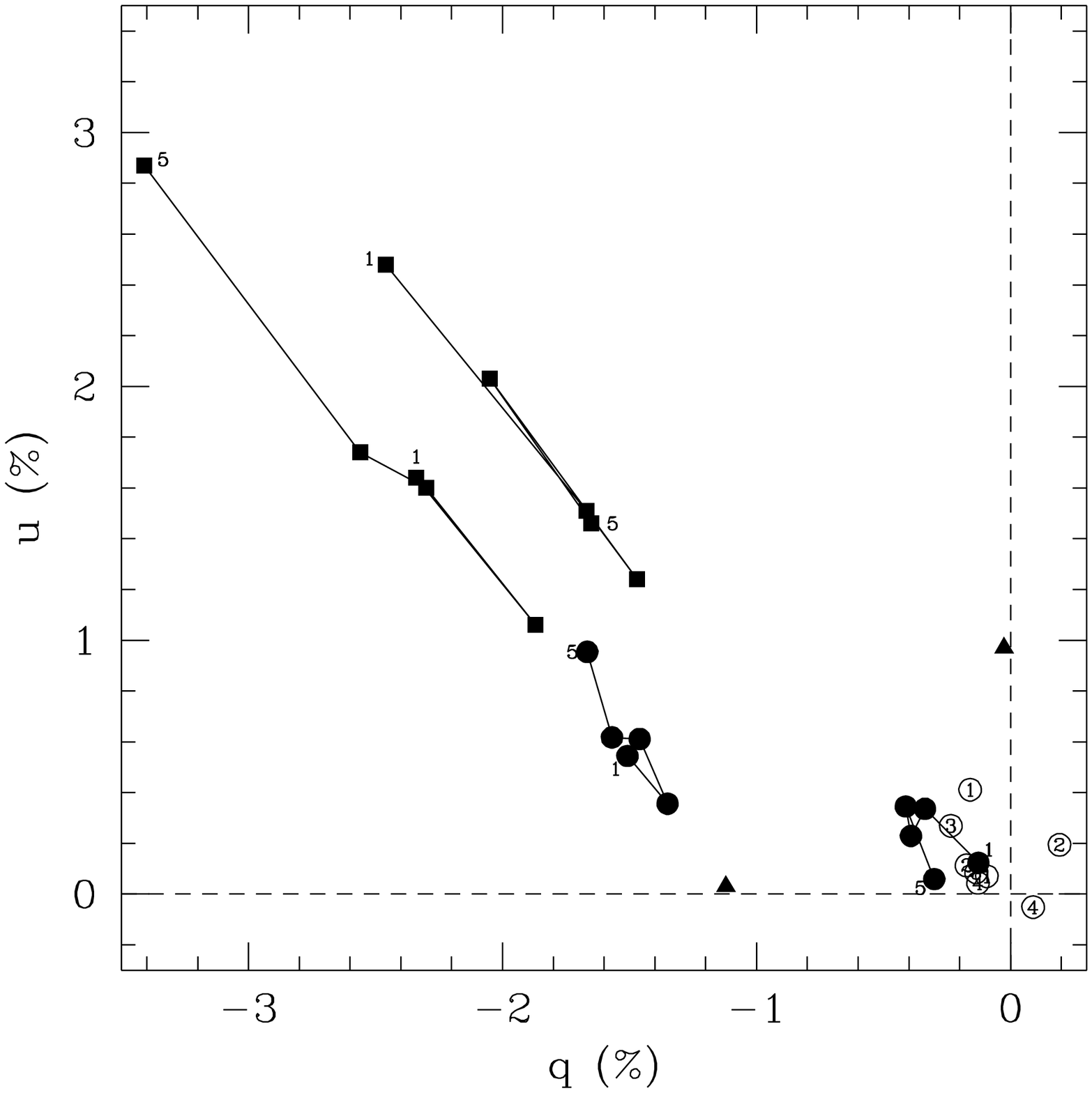}
\caption{Observed continuum polarizations (solid dots) and observed broad \ha~wing polarizations 
(solid squares) from Table \ref{olog} of NGC 2110 (blue) and NGC 5252 (green) in the $q$-$u$ plane. 
Numbers 1 and 5 correspond to the first and last epochs of 
observations, respectively. The solid lines connect adjacent epochs. Also plotted in open circles are 
the polarizations of the Galactic ISpol probes from Table \ref{isp} for NGC 2110 (blue) and NGC 5252 
(green). Numbers 1-4 denote the star numbers in Table \ref{isp}. Solid triangles denote the narrow-line
polarizations for NGC 2110 (blue) and NGC 5252 (green), as determined from the 
\oiii~$\lambda$$\lambda$4959, 5007 
lines. 
\label{contpol}}
\end{figure*}

Figure \ref{contpol} shows the observed continuum polarizations in the $q$-$u$ plane for NGC 2110 and 
NGC 5252 from Table \ref{olog}, along with the polarizations of the ISpol probes from Table \ref{isp}. 
As Table \ref{isp} and Figure \ref{contpol} show, the ISpol observed from the probes for NGC 5252 are 
relatively small and fairly consistent with each other, and we adopt the average result of the top two  
highest polarized probes ($P$ = 0.18\%, $\theta$ = 74\arcdeg) as the ISpol toward NGC 5252. 
For NGC 2110, the results from the ISpol probes are more difficult to interpret since the observed 
polarizations from the four selected stars are more ``scattered'' or discrepant from each other, 
and their magnitudes are comparable to those observed for the continuum of NGC 2110. Therefore, more 
care needs to be applied when selecting the most appropriate ISpol. We can apply two criteria to help us
make this selection. One is that the ``best'' ISpol would preserve the perpendicular relationship 
between the polarization P.A. and the well-determined radio and ionization cone axes of NGC 2110. 
The second is that an appropriate ISpol correction should maintain a similar polarization P.A. between 
the continuum and broad \ha~line \citep{t95}. The ISpol correction that satisfies both of these tests 
is one from the probe PPM 188568 (star \#4, $P$ = 0.11\%, $\theta$ = 165\arcdeg), and we adopt this as 
the most representative Galactic ISpol toward NGC 2110. This is also the most conservative estimate 
of the ISpol, as it has the smallest magnitude of the four stars and does not significantly rotate the 
observed $\theta$. For comparison, \citet{m07} used $P$ = 0.33\%, $\theta$ = 34.5\arcdeg~as the ISpol. 
The correction of ISpol was made by fitting a Serkowski curve (Serkowski, Mathewson, \& Ford 1975) to 
the adopted ISpol and subtracting it
from the observed $q(\lambda)$ and $u(\lambda)$ of the galaxies.
It is worth noting that neither of the Galactic ISpol adopted here for NGC 2110 and NGC 5252 made a 
significant modification to the observed polarizations. Their corrections are minor, and the 
conclusions reached in this study are not sensitive to the adopted ISpol. 

\begin{figure}
\plotone{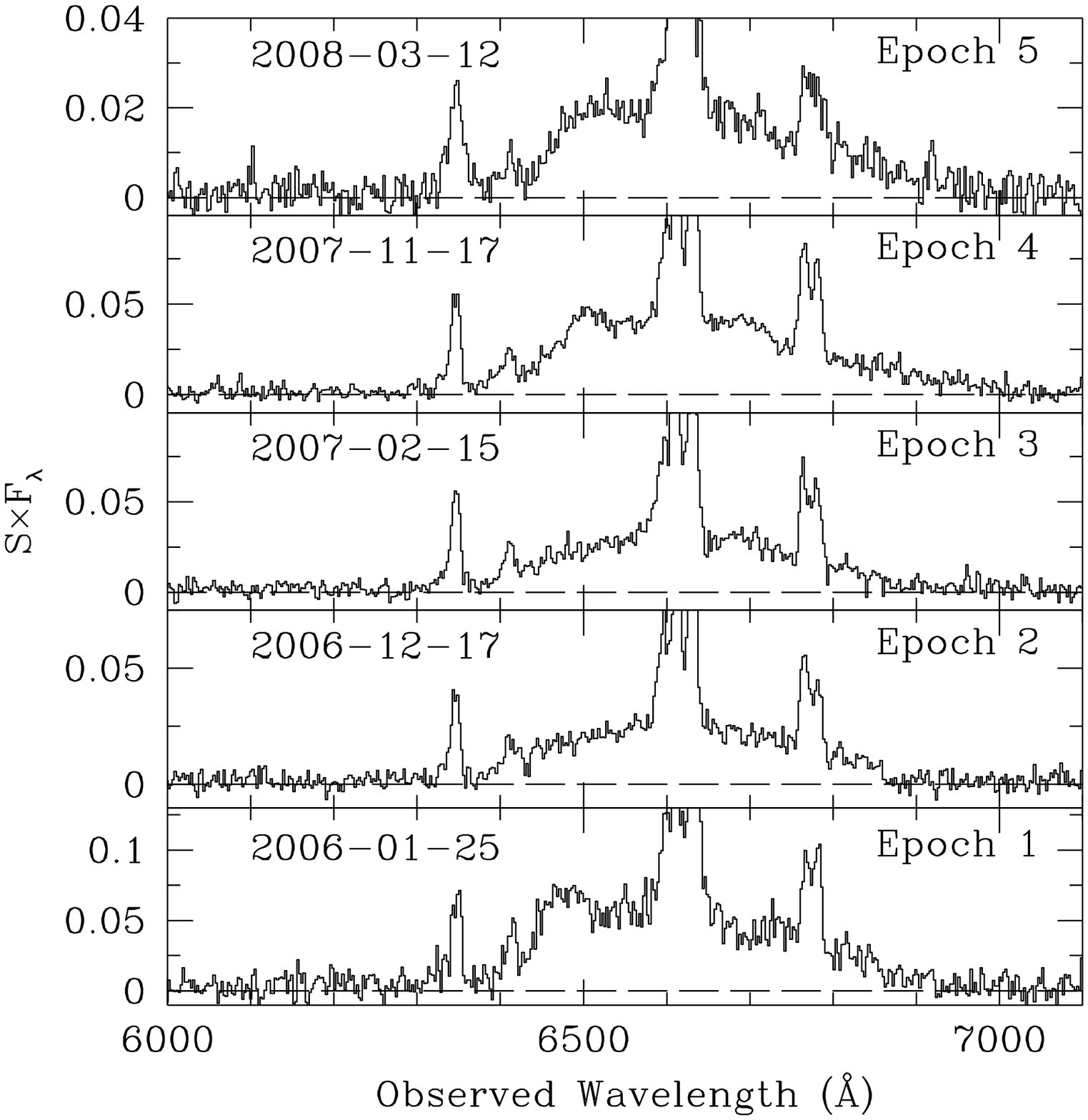}
\caption{Continuum-subtracted \ha~profiles in polarized flux of NGC 2110 for all five epochs of 
observations. The flux scales are in units of 10$^{-15}$ \flxu. Note the dramatic variation in strength
and profile shape of the polarized broad \ha~emission line.
\label{n2110ha}}
\end{figure}

\subsection {Continuum, Narrow-line, and Broad \ha~Polarizations} \label{blpol}

Figure \ref{contpol} also displays the polarizations in the broad \ha~wing. As can be seen, the 
observed polarizations are much higher there than in the continuum. This is the result of 
lessened dilution by the underlying host galaxy in the \ha~emission because of the higher flux in
the line compared to the continuum. Significant variations in polarizations in both the continuum and 
broad \ha~lines are observed, with somewhat higher magnitude in the line. For example, for 
NGC 2110 in epochs 2$-$4, when the continuum polarization remained relatively unchanged, the \ha~wing
polarization went from 1.9\% to 2.9\%, a change by a factor of $\sim$1.5. Similarly, in NGC 5252 the
broad \ha~wing $P$ varied by a factor of 2 over all epochs, while the continuum $P$ remained within a 
factor of $\sim$ 1.3 of each other.  
Close examination of Figure \ref{contpol} also shows that the observed polarization P.A. 
$\theta$ seems to significantly vary over the observed epochs.  In addition, there appears to be 
a slight rotation of $\sim$ 6\arcdeg~between the continuum and broad \ha~$\theta$; i.e., the 
polarization changes are not along a radial vector on the $q$-$u$ plot, as in NGC 2110. The direction and 
small magnitude of the inferred Galactic ISpol cannot account for this rotation, implying that there 
is another polarized component at play. 
This third polarized component (besides the scattered AGN continuum and broad 
\ha~line) could be the substantial (see below) host galaxy starlight which has traditionally been 
assumed to be unpolarized, but may be slightly polarized in these cases, perhaps due to dichroic 
absorption within the AGN host itself. One piece of evidence in favor of this interpretation is that 
the $\theta$ variation observed in the continuum ($\sim$ 7\arcdeg-17\arcdeg) is greater than that in 
\ha~($\sim$~few degrees), due to the greater relative contribution of starlight in the continuum 
than in the emission line. Consistent with this, $\theta_c$ during ``high'' continuum state appears 
to agree well with $\theta_{H\alpha}$ in ``low'' broad \ha~state. Compare, for example, 
$\theta_{H\alpha}$(epoch 3) with $\theta_c$(epoch 4) of NGC 2110, and $\theta_{H\alpha}$(epoch 2) with
$\theta_c$(epoch 5) of NGC 5252. If the third polarized component comes from the ISpol in the host 
galaxy itself, it would induce a polarization in the narrow lines. Indeed, narrow emission lines are 
clearly seen in the polarized flux spectra. We measure the narrow-line polarization from the  
\oiii~$\lambda$$\lambda$4959, 5007 emission lines and plot them as solid triangles in 
Figure \ref{contpol}.
The narrow-line polarization is $P_{NL}$ = 1.1\%, $\theta_{NL}$ = 89\arcdeg~for NGC 5252, and 
$P_{NL}$ = 0.98\%, $\theta_{NL}$ = 46\arcdeg~for NGC 2110. It is easy to see from the figure that a 
correction for the narrow-line polarization in NGC 5252 would result in a radial change in 
polarization P.A. on the $q$-$u$ plane for both the continuum and broad line at a P.A. of $\sim$ 68\arcdeg, 
thus strengthening the host ISpol origin. On the other hand, correcting for the 
narrow-line polarization in NGC 2110 does not preserve such a relationship, and we believe its 
narrow-line polarization probably arises from scattering in the narrow-line region (NLR) itself.    

To within only a few degrees, the polarization P.A.s after correction for ISpol are closely 
perpendicular to the radio and ionization cone axes in both galaxies. For NGC 5252, polarization 
$\theta$ is $\sim$ 70\arcdeg, compared to a P.A. of $\sim$ 345\arcdeg~for the radio and emission-line 
cone axis \citep{wt94}. For NGC 2110, the polarization $\theta$ $\sim$ 70\arcdeg, orthogonal to the 
radio and emission-line structure P.A. of $\approx$ 150\arcdeg-170\arcdeg~\citep{mul94,mid04}. In both 
cases, the polarization P.A. is essentially the same between the continuum and the broad \ha~emission 
line to within the uncertainty.  

The observed variation in the broad polarized \ha~emission can be attributed directly to variations
in the emitting line flux, as the intrinsic broad-line polarization does not appear to change over time.
The {\it intrinsic} broad \ha~polarizations derived by dividing the continuum-subtracted broad-line 
flux in the 
polarized flux spectrum over its counterpart in the total flux spectrum yield $p$(\ha) $\sim 10$\% for 
NGC 5252 and $p$(\ha) $\sim 20$\% for NGC 2110 over the epochs observed. The uncertainty of the 
measurement is $\sim \pm$5\%. The observed variation in the continuum and broad-line polarizations is 
consistent with a changing continuum and broad-line fluxes over a non-changing underlying stellar 
continuum of the host galaxy, as the polarization P.A. remains approximately constant in all epochs. 
An independent 
estimate of the galaxy fraction using the elliptical galaxies NGC 821 and NGC 6702 (see Tran 1995a) 
as templates indicates that the galaxy fraction $f_g$ is $\sim$ 0.85 and 0.83--0.95 for NGC 5252 and  
NGC 2110, respectively. This implies that the intrinsic, galaxy-dilution corrected continuum 
polarization (assuming an unpolarized galaxy component) is $\sim$ 10\% for NGC 5252 and $\sim$ 5\% 
for NGC 2110. This agrees with the intrinsic broad 
\ha~polarization of $\sim$ 10\% in NGC 5252 derived above, but falls well short of the estimated 
20\% for NGC 2110, suggesting that another significant source of unpolarized light may be 
present \citep{t95b,sm02}. For both objects, the galaxy corrected $p(\lambda)$ is relatively flat, 
suggesting that the diluting source of light is nearly independent with wavelength.  
We note in passing that, along with the LINER-like emission-line ratios displayed by 
NGC 2110 \citep{m07} and NGC 5252 \citep{gon98}, a strong starlight-dominated optical continuum is 
also an important secondary characteristic of the double-peaked emission-line AGNs \citep{e04}.

\subsection{Variability of the Polarized Broad \ha~Profiles}

The exceptional discovery of this study is that the polarized broad lines from the nuclei of NGC 2110
and NGC 5252 have been observed to vary with time, which is quite unique and unprecedented about 
these objects. 
Figure \ref{n2110ha} shows the continuum-subtracted profiles of \ha~in polarized flux for NGC 2110.  
In epoch 1, note the clear blue peak in the polarized \ha~emission line, and the correspondingly
high polarization in the blue wing (see Figure \ref{n2110}, Table \ref{olog}). This profile most closely 
resembles (as it was closest in time) that observed in 2005 December by \citet{m07}. As discussed
by \citet{m07}, this line profile is very similar to that of the prototypical DPE 
Arp102b. By 2006 December and 2007 February (epochs 2 and 3), this blue peak has disappeared, and the
polarization there is also much smaller.  By 2007 November (epoch 4), the blue peak is starting to 
come back, with a correspondingly higher polarization there. 

\begin{figure}
\plotone{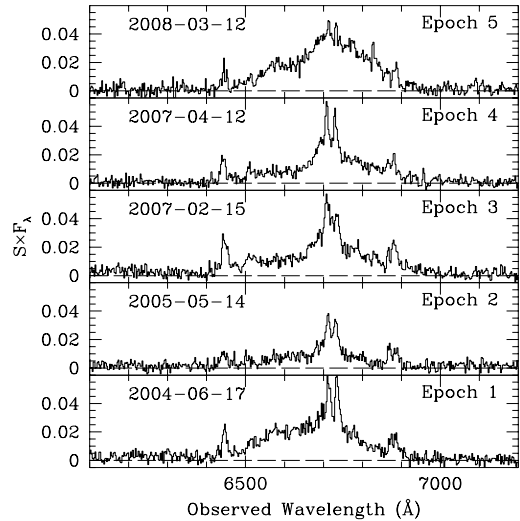}
\caption{Same as Figure \ref{n2110ha} but for NGC 5252. Note the dramatic variation in strength and more
subtle change in profile shape of the polarized broad \ha~emission line.
\label{n5252ha}}
\end{figure}

Similar variations are also observed in NGC 5252, as shown in Figure \ref{n5252ha}. Here we show the 
continuum-subtracted polarized flux spectra around \ha~for all five epochs. It can be seen 
that the polarized broad-line intensity was fairly prominent in 2004 June (epoch 1), became 
significantly weaker almost a year later in 2005 May (epoch 2), got stronger again in 
2007 February and April (epochs 3 and 4), then became strongest in 2008 March (epoch 5). 
One notable difference 
compared to NGC 2110 is that while the polarized broad \ha~flux varied dramatically, this does not seem
to be accompanied by a great change in the shape of the profile. The double-peak nature also 
appears less pronounced here compared to NGC 2110.  

Clearly, the polarized broad \ha~is changing both in shape and intensity on timescales of months. 
Over the smallest time interval between two adjacent epochs, which is about 2 months for both 
NGC 2110 and NGC 5252, changes are generally not seen, but definite changes are clearly observed on 
timescales $\lesssim$~1 yr. Such dramatic variations on such short timescales in both the degree of 
polarization and profiles of the {\it polarized} broad emission lines have never been observed before 
in other classical hidden broad lines of S2s. 
In general, variability is not expected in reflected, polarized light, since the scattering process 
tends to smear out any intrinsic variations.  
Based on the elliptical disk fit to the broad \ha~line of NGC 2110,  \citet{m07}  found that the inner radius 
of the line-emitting disk is  $\sim$ 200 gravitational radii ($r_g \equiv GM_{BH}/c^2$, where $M_{BH}$ is the 
black hole mass). This is rather small compared to many DPEs \citep[see e.g., ][]{st03}, 
suggesting that any changes taking place in this part of the disk could lead to rapid variations. Continued 
regular monitoring of objects like NGC 2110 would provide better constraints on the timescales of the variability.  

The observed polarization variability can be summarized as follows.
In the continuum, the polarization remains little changed throughout the period being monitored, 
with $P$ being $\sim$ 1.5\% for NGC 5252 and $\sim$ 0.5\% for NGC 2110.
However, in the broad \ha~emission line, both the observed polarization and polarized flux vary 
substantially with time. Interestingly, there is {\it no} corresponding large changes in the 
polarization P.A. Furthermore, the \ha~polarization P.A. is the essentially same as in the 
continuum, suggesting that they have similar scattering geometry. 
Finally, the polarization P.A. is closely perpendicular to the P.A. of the extended ionization cones and
radio axes for both galaxies, confirming that scattering is the cause of polarization. 

\section{Discussion} \label{disc}

\subsection{Variability of the Scattered Line Profile} \label{var}

The observed high degree of perpendicularity of the polarized P.A. to the radio and ionization cone 
axes in both NGC 5252 and NGC 2110 and the high intrinsic polarization ($\sim 10\%-20\%$) of the broad 
\ha~emission line clearly establish that scattering is the only viable mechanism producing the 
observed polarization.  

The most interesting finding of the present study is the discovery that the broad double-peaked 
\ha~emission line from the nucleus of NGC 2110 displays dramatic changes in profile 
and intensity in polarized flux spectra on timescales $\lesssim$ 1 yr. Dramatic variation in the 
polarized flux intensity of \ha~is also seen NGC 5252, but detailed profile variation of its broad line
is less pronounced than in NGC 2110. Such behavior of the 
polarized broad lines from type 2 AGNs is entirely unexpected and has never been reported previously 
in other HBLR S2s, as far as we are aware \citep{t95,t01}. 
Temporal and structural polarization variability is common in type 1 AGNs, such 
as Seyfert 1s and broad-line radio galaxies (see e.g., Goodrich \& Miller 1994; Martel 1998; 
Cohen \etal~1999; Smith \etal~2005), but these variations are thought to be due to near-field 
scattering in an equatorial disk just outside the BLR \citep{sm05}. Such process cannot account for 
the behavior that is observed in NGC 2110 and NGC 5252 since the BLR and equatorial scattering region 
are both entirely obscured in these objects. 

Nor can the rapid polarization flux variability be explained by the ``light-echo'' or ``search-light'' 
effect, in which differentially redirected light in a clumpy medium mimics the temporal variation of 
the polarization signal. Such an effect may be able to reproduce changes in the degree of polarization 
and polarized flux level, but cannot explain the observed changes in the structures of the 
emission-line profiles.  This intrinsic variability of the polarized emission-line profile, coupled 
with the non-changing polarization P.A. of the emission line over time, suggests that the 
variations are due to changes in structures of the line-emitting region itself, {\it not} the 
scattering medium. 

Profile variability of the broad \ha~emission line has been known to be a very common property of 
the ``normal'', directly viewed DPEs \citep{e04}. If emitted from an 
accretion disk, the relevant thermal and dynamical timescales, summarized in \cite{e04}, range from 
days to hundreds of years depending on the mass of the accreting black hole. The black hole mass of 
NGC 5252 has been measured by \citet{cap05} to be $\sim 10^9$~\msun. \citet{m07} estimated the black 
hole mass within NGC 2110 to be $\sim 2 \times 10^8$~\msun~using its observed stellar velocity 
dispersion and the \mbh-$\sigma$~relation.

We can then estimate and compare these timescales for NGC 5252 and NGC 2110. The observed polarized 
profile variations in NGC 5252 and NGC 2110 cannot be due to thermal or sound-crossing phenomena 
because the timescales involved are too long -- of order tens to hundreds of years. They are more 
consistent with the dynamical timescale for NGC 2110 ($\sim$ 12 months) or light-crossing time in 
NGC 5252 ($\sim$ 2 months). This is consistent with the fact that the polarized \ha~variability in 
NGC 2110 is accompanied by significant profile variation, while that in NGC 5252 generally does not, 
suggesting that the observed polarized line flux variation in the latter might simply be due to 
response of the line emitting region to reverberation of a changing continuum flux. 

\subsection{Nature of the Scatterers} \label{scat}

Our observations can put several important constraints on the properties of the scattering medium.
The polarization after correction for starlight dilution, and the observed polarized flux spectra are 
relatively flat and not significantly bluened as expected from Rayleigh scattering by normal dust 
grains. This is consistent with electron scattering, although dust scattering by fine grains with the 
right properties or in a clumpy medium can also produce wavelength-independent polarized light 
\citep{k95, v01}. For simplicity, we shall assume electron scattering as the main scattering mechanism 
for the rest of our discussion. 

As previous mentioned by \citet{m07} and seen in our data, the shape of the scattered broad \ha~is 
remarkably similar to other normal DPEs, especially Arp 102B, suggesting little modification or 
smearing of the line profile in the scattering process. Although the exact amount of broadening is 
difficult to determine because the line is so broad, this places a constraint on the temperature 
of the scattering electrons to be \te~$\lesssim$ $10^6$ K \citep{mgm91}. 
Discussion in the previous section also suggests that the scattered flux must be dominated by a few 
individual discrete clouds in a clumpy medium instead of a cone largely filled with material, as 
typically assumed in S2s (i.e., Code \& Whitney 1995). 
The observed variability timescales constrain the size of the scattering clouds to be very compact,
$\lesssim$ 1 lt-yr in size. Because we must be able to observe the scattered light, and 
because the polarization P.A. is perfectly perpendicular to the radio and bicone axis, these scatterers
must be distributed along the polar direction outside of the obscuring torus. 

The scattered fraction of light from a spherical cloud located at a distance $d$ from the nucleus with 
radius $r$ can be approximated as
\begin{equation}
f \equiv \frac {L_{sc}}{L_{in}} \approx \sigma_T n_e 2r \Delta \Omega = 2 \sigma_T n_e r^3 /d^2, 
\end{equation}
where we have assumed the probability of scattering to our line of sight is near unity, 
\lsc~is the scattered luminosity, \lin~is the intrinsic or incident luminosity of the obscured 
nucleus, $\sigma_T = 6.65 \times 10^{-25}$ cm$^{2}$ is the Thomson scattering cross section, 
\nee~is the mean electron density, and $\Delta \Omega \approx \frac {1}{4} (2r/d)^2 = r^2/d^2$ is 
the fraction of solid angle subtended by the scattering cloud. 
If we assume that the scattering is optically thin to electrons, the optical depth 
$\tau_e = \sigma_T n_e 2r \lesssim 1$, and $2r$ is known from our observations to be $\sim$ 1 lt-yr, 
it follows that the required electron density is \nee~$\sim 10^7$ \cc~and  
\begin{equation}
d \lesssim 10 f_\%^{-1/2} n_{e7}^{1/2} r_{ly}^{3/2} ~{\rm pc}, 
\end{equation}
where $f_\%$ is the scattered fraction in percent, $n_{e7}$ is the electron density in units 
of $10^7$ \cc, and $r_{ly}$ is the radius of the scattering cloud in light years.   

The fraction $f$ can be estimated from the observed luminosity of the scattered broad \ha~flux in
NGC 2110 and NGC 5252 and comparing them to the directly viewed \ha~luminosity of the normal DPEs. 
The observed scattered broad \ha~flux in NGC 2110, and NGC 5252 are typically $\sim 1.5 \times 10^{-14}$
and $9.2 \times 10^{-15}$ \flxu, respectively. Correcting these values for the $\sim$ 20\% and 
$\sim$ 10\% polarization of the broad line, respectively, this corresponds to a scattered broad 
\ha~luminosity of $\sim 10^{40} - 10^{41}$ \lumu~for these two hidden DPEs (HDPEs). 
The broad double-peaked \ha~luminosity in a sample of directly viewed DPEs is typically 
$\sim 10^{42} - 10^{43}$ \lumu~\citep{eh03}, about 2 orders of magnitude higher. Assuming that 
NGC 2110 and NGC 5252 are the exact type-2 counterparts to these directly viewed DPEs, this implies 
that the scattered faction $f \sim 1\%$, and Equation (2) indicates that the scattering clouds need 
to be of order $\lesssim$ 10 pc from the nucleus, placing them just outside the obscuring torus 
(see e.g., \citet{ra09}, and references therein), and between the BLR and NLR. This is also consistent with
\citet{mas09}, who recently constrain the outer radius of the obscuring torus in NGC 2110 to $< 8$ pc.  
The electron density 
derived is also consistent with this location of the scatterers, lying just between typical values for
the BLR ($\sim 10^{10}$ \cc) and NLR \citep[$\sim 10^{4}$ \cc;][]{o93}. With density 
\nee~$\sim 10^7$ \cc~the required ionized gas mass for each light-year-wide scattering cloud is 
$\lesssim 10^3$ \msun, assuming a filling factor of unity within the cloud, which may be an 
overestimate.  The derived distance of $\lesssim$ 10 pc for the scattering clouds is also consistent 
with our assumption of electrons as the dominant scatterers, as dust may have more difficulty
surviving in the harsh environment in close proximity to the active nucleus.  

Note that such a scattering region is much more compact and close-in to the nucleus than those 
previously envisioned for the classical HBLR S2s, where the scattering region is thought to be the 
size of the extended NLR, or of order $\sim$ $10^2 - 10^3$ pc. 
Scattering from such an extended scattering NLR may still take place in NGC 5252 and NGC 2110, but our 
current observations dictate that reflection from these very compact, close-in scattering clouds 
must dominate the polarized light from these nuclei.  In order for these scattering clouds 
to not ``smear'' out the variability, there must not be many of them along each ray, 
perhaps numbering $\lesssim$ 10.  
Similar compact scattering region has been proposed by \citet{gal05} to explain the 
polarization behavior of the Seyfert 1 galaxy Mrk 231. Based on ground-based and 
$Hubble~Space~Telescope~(HST)$ polarization 
observations, the lack of any spatially extended polarization structures and the presence of polarization 
P.A. structure across the broad emission lines place the dominant scatterers in Mrk 231 to within 
20 pc of the nucleus in a polar lobe distribution. This could be analogous to the compact scattering 
region indicated for NGC 2110 and NGC 5252.

We now speculate as to what this compact scattering region might be. 
We consider three possibilities: (1) line emitting gas ``ejectiles'' from the nucleus, 
(2) radio hot spots or material entrained in the base of the jets, or 
(3) material from the outskirts of the obscuring torus itself.
It is noteworthy that all three possibilities discussed here have one common feature: 
they all involve AGN feedback -- winds or outflow of material driven by the central engine. 

The ejection or bipolar outflow model \citep{zbs90} that was proposed to explain 
the double-peaked broad emission lines in normal DPEs could provide the natural source of material 
near the nucleus to scatter the obscured nuclear continuum and broad lines. This scenario seems 
attractive as it could provide the source for both the BLR and scattering gas clouds, and at the same 
time explain the double-peaked nature of these sources. It is unclear, however, whether the same gas 
clouds can serve for both purposes, as they must remain compact ($r \lesssim 1$ lt-yr) after having been 
driven out to $\sim$ 10 pc in the outflow. There may also be some difficulty for the scatterers to see
both polar ejecta at this relatively close distance from the central source. 

The radio jet picture is also a plausible candidate since it provides a natural explanation for the 
preference of DPEs in radio-loud AGNs, in which $\sim$ 20\% of DPEs are found, compared to only 
$\sim$ 3\% in the general AGN population \citep{st03}. Interaction of the radio jets with material 
immediately surrounding the central source could produce ionized gas or ``hot spots'' that could serve 
as the scattering medium. Since these jets are thought to be highly collimated, it is probably not 
difficult to produce very compact scattering plasma clouds, as required. Very Large Array (VLA) imaging 
of NGC 2110 by \citet{nag99} indeed revealed a radio jet extending $\sim$ 400 pc from a central core. 
Very high resolution Very Long Baseline Array (VLBA) imaging by \citet{mun00} capable of resolving subpc-scale details, 
subsequently showed some slightly resolved emission $\lesssim$ 1 lt-yr in size extending $\sim$ 1 lt-yr
from the nuclear core in the same direction as the hundreds-pc scale jet. Although the separation  
from the central engine may be too small, it is conceivable that these extended ``knots'' are 
associated with a compact scatterer. 

Finally, the hypothesized obscuring torus central to the AGN unification model could provide readily 
available material for the scattering medium. Perhaps the individual torus clouds in the clumpy 
torus model of \citet{es06} and \citet{nen08} could themselves serve as the scattering mirrors. 
As discussed by \citet{nen08}, the obscuring clouds could be either dusty or dust-free, and because 
the appearance of a type 1 or type 2 AGN is probabilistic, dependent partly on the number of clouds 
and not solely on the viewing angle, these clouds could play a role as both an obscuring source and 
the ``polar'' scattering region, as required. It is especially compelling to note that the physical
properties of these clumpy torus clouds \citep[i.e., \nee~$\sim 10^7$, $r \lesssim 1$ lt-yr, $m \sim$ $10^2$ 
\msun, adjusted for $d \lesssim 10$ pc and black hole masses of NGC 2110 and NGC 5252;][]{es06}
are strikingly similar to those constrained for the scattering clumps in NGC 2110 and NGC 5252.

\subsection{Implications for the Unified Model and Double-peaked Emission Line AGNs} \label{dpe}

\begin{figure}
\plotone{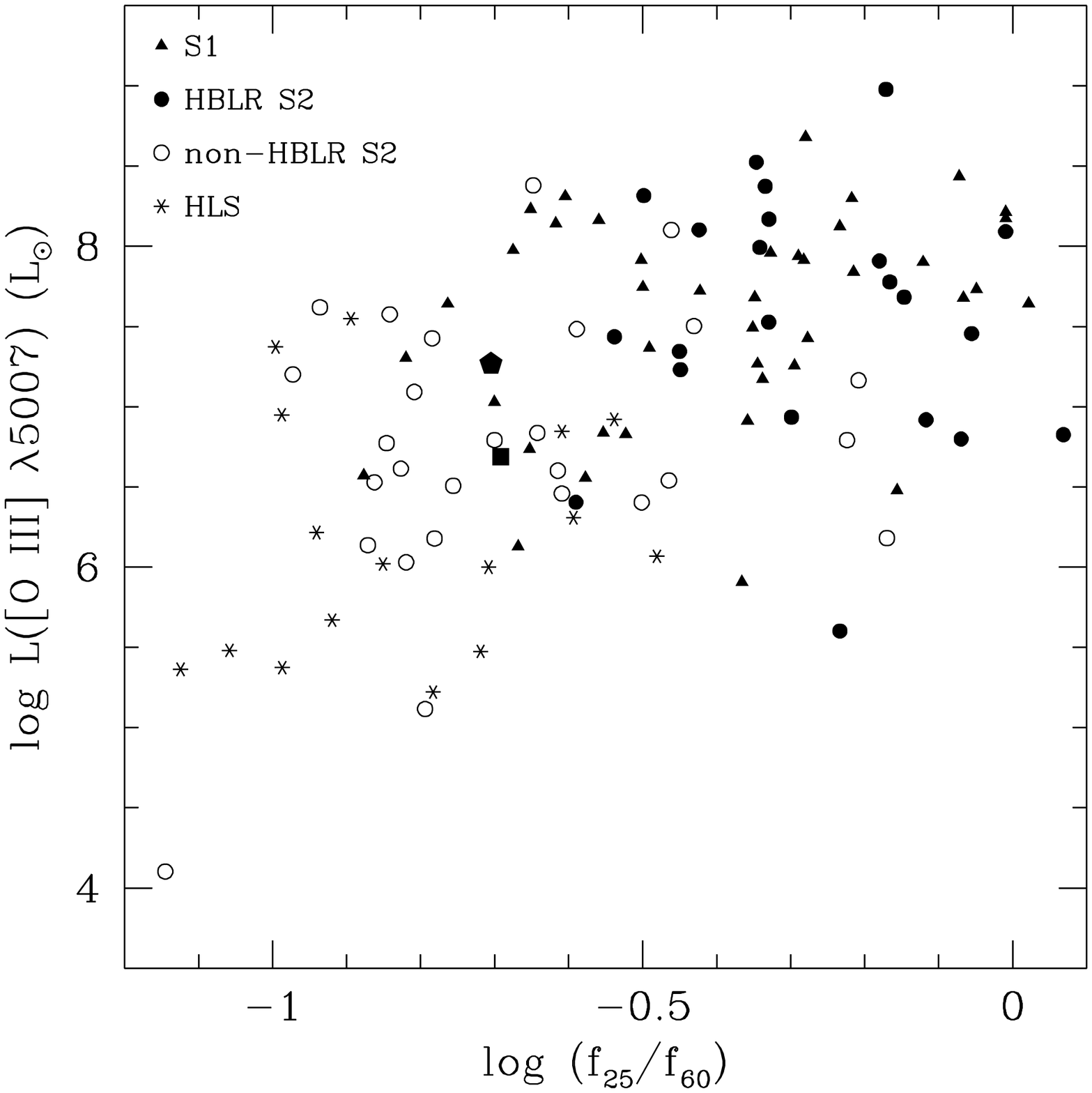}
\caption{Positions of NGC 2110 and NGC 5252 on the \oiii~\wave 5007 luminosity vs. IR color 
\firr~diagram, reproduced from \citet{t03}. ``HLS'' represents H II region galaxies, LINERs and 
starburst galaxies. NGC 2110 and NGC 5252 are represented by the black square and pentagon, respectively. 
We used the $Infrared~Space~Observatory~(ISO)$ \ftv~and \fst~fluxes from \citet{pa03} for NGC 5252 and 
$IRAS$ fluxes for NGC 2110. \oiii~\wave 5007 fluxes have been measured from our current spectra. Both 
galaxies lie in the non-HBLR region of the diagram, suggesting that their HBLR detections may be owed to their polarized broad-line variability. 
\label{trandia}}
\end{figure}

The discovery of two HDPEs in Seyfert 2s suggests that such objects
may be common. Other similar hidden extremely broad lines previously seen are probably those observed 
in Cygnus A \citep{o97} with polarized \ha~FWHM of $\sim$ 26,000 \kms, and 2MASSI J130005.3+163214 
\citep{sm02} with polarized \ha~FWHM $\sim$ 18,000 \kms. 
On the other hand, that they can vary even in polarized flux means that many may also
escape detection if not followed systematically by spectropolarimetric observations. The variability 
that these HDPEs exhibit may also partly explain why about half of the Seyfert 2s do not show any 
detectable HBLRs in spectropolarimetric surveys, inconsistent with the simplest unification scheme 
\citep[e.g.,][]{t01,t03}. \citet{t03} found that the HBLR Seyfert 2s tend to lie in the ``hotter'',
``stronger'' part of the \oiii~\wave 5007 luminosity versus IR color \firr~diagram (his Figure 9), while 
the non-HBLRs in the ``cooler'', ``weaker'' area. Figure \ref{trandia} shows that both NGC 2110 and 
NGC 5252 lie in the non-HBLR region of this diagram, suggesting that if it had not been for their 
variability, their HBLRs might not have been detected. Indeed previous spectropolarimetric 
observations of these sources at Lick Observatory did not reveal anything of interest \citep{kay94}) 
in polarized flux, and \citet{y96} reported only a marginal detection with significant uncertainty of 
polarized broad \ha~in NGC 5252. However, this could also be due to the inadequate depth of the 
observations with smaller telescopes. Also, \citet{kay94} observations only extended to \hb~and did 
not include \ha, and broad polarized \hb~is difficult to detect even in the present observations.

\citet{eh03} present a good assessment of the various models for the DPEs in light of their 
comprehensive survey of radio-loud AGNs, and conclude that overall, emission from an accretion disk 
appears to be best at explaining all the various properties of the DPEs.  They disfavor, but do not rule out,
other models, such as binary BLRs, bipolar outflows, and anisotropically illuminated spherical 
BLRs. While our observations do not clearly favor one model over others, it does suggest that bipolar 
outflows are still viable and less unlikely.
If the scatterers are 
``spent'' BLR clouds ejected from the nucleus, then they may naturally serve as the same 
material responsible for the double-peaked emission lines. Thus, the bipolar outflow model does 
provide a simpler picture for the two necessary ingredients of the HDPEs: double-peaked broad-line 
emitting gas, and compact scattering material close to the nucleus. Also, the well-defined bicone of 
ionized gas in NGC 5252 and NGC 2110 could serve as a natural extension of the bipolar outflow from 
the obscured nuclei.  

If the scattering clouds are radio hot spots or material entrained in the jet 
(see discussion in Section \ref{scat}) then they necessarily must lie close to the radio axis, which 
presumably is the same as the axis of the accretion disk itself. 
This may present a difficulty for the accretion disk model. 
An elliptical accretion disk fit 
to the polarized double-peaked \ha~line profile by \citet{m07} indicates that the scatterers view 
the disk at an inclination angle of $\sim$ 30\arcdeg. This best-fit angle may be less secure 
than those resulting from the modeling of other directly viewed DPEs, because the exact contribution 
of light to this reflected broad-line profile is unknown. 
Nevertheless, taken at face value, such viewing orientation seems uncomfortably far from the disk 
axis, since the radio half opening angles in Seyfert galaxies are generally well within 
$\sim$ 15\arcdeg~\citep{wt94}. For NGC 2110 in particular, the ionization cone is actually more 
``jet-like'' \citep{mul94}, further suggesting that any scattering material from the jet should lie 
close to the pole. 

If however, the scattering clouds come from the clumpy obscuring torus, then there is no restriction 
on the orientation of the scatterers relative to the radio axis as long as they lie within the 
ionization cones. In this case, the accretion disk model may be preferred, based on the very extended 
bicone morphology of the ionization structures in these galaxies, indicating that our viewing angle 
must be fairly large, most likely $\gtrsim$ 50\arcdeg~\citep{t00}. The derived high intrinsic 
broad-line polarizations of $\sim$ 10\%-20\% (see Section \ref{blpol}) indicate that the scattering angle is 
$\sim$ 30\arcdeg-40\arcdeg, based on the models by \citet{cw95} of an externally illuminated spherical 
electron-scattering blob. This suggests that our viewing angle is $\sim 30 + 35 = 65\arcdeg$ for the 
accretion model, more consistent with the observed extended ionization cone morphology than the 
inclination of $\lesssim 15 + 35 = 50\arcdeg$ inferred for the bipolar outflow/scattering radio jet 
model. 

\section{Conclusions} \label{con}

From these observations we can draw the following conclusions.
The fact that we detect any variations in polarized broad line at all indicates that the 
scatterers are physically very compact, with size scales of order 1 lt-yr, similar to the 
dynamical timescales of DPEs. 
This suggests that the scattering is primarily done by a few discrete clouds 
rather than in a filled cone with a large filling factor, as previously assumed for S2s.
Second, because the continuum polarization and P.A. remain relatively little changed, with the P.A. 
being the same as in the broad emission line, these variations are most likely due to changes in the 
line emitting flux, perhaps because of changes in the structure of the line emitting region, 
and {\it not} the scattering medium. 

With the reasonable assumption of electron scattering, we constrain the size, location, temperature, and
density of the ionized gas clouds responsible for the polarization. We find that the scattering 
clouds are $\lesssim$ 1 lt-yr in size, confined to $\lesssim$ 10 pc of the nucleus, having densities 
$\sim 10^7$ \cc~and temperatures $\lesssim 10^6$ K. This is probably similar to the compact scattering 
region proposed to exist in the Seyfert 1 galaxy Mrk 231 by \citet{gal05}. We speculate that the 
scattering region could arise from gas ``ejectiles'' from the bipolar outflow, hot spots or material 
entrained in the base of the radio jets, or clumpy clouds from the outskirts of the obscuring torus 
itself. The derived physical properties of these scattering clouds are consistent with those of the
clumpy torus clouds of \citet{es06} and \citet{nen08}.

Finally, because they share many characteristics similar to the DPEs, NGC 5252 and 
NGC 2110 are the type-2 or hidden counterparts of this class of objects, which has not been found 
until recently.  Continued spectropolarimetric monitoring of these objects would be very valuable 
in elucidating both the nature of the DPEs and their connection to the general AGN population as a whole.

\acknowledgments
This research would not be possible without the foresight and generosity of the W. M. Keck Observatory 
directors, who have shared their observing time with the ``Team Keck'' scientific staff at the 
Observatory. I am very grateful to F. Chaffee and T. Armandroff for their continued support. 
Part of this research was carried out while HDT enjoyed a very fruitful scientific visit at ESO. 
I thank R. Fosbury, J. Vernet and C. De Breuck for making this visit possible, and for 
many illuminating discussions. I acknowledge insightful discussions with D. Baade, 
E. Emsellem, S. Veilleux, and thank R. W. Goodrich, P. Ogle, and R. Fosbury for constructive 
comments on an early draft of the paper.  This paper benefited from several useful comments from 
an anonymous referee, to whom I am grateful.
The data presented herein were obtained at the W. M. Keck Observatory, which is operated as a 
scientific partnership among the California Institute of Technology, the University of California and 
the National Aeronautics and Space Administration. The Observatory was made possible by the generous 
financial support of the W. M. Keck Foundation. I wish to recognize and acknowledge the very 
significant cultural role and reverence that the summit of Mauna Kea has always had with in the 
indigenous Hawaiian community. We are most fortunate to have the opportunity to conduct observations 
from this mountain. This research has made use of the NASA/IPAC Extragalactic Database (NED), which 
is operated by the Jet Propulsion Laboratory, California Institute of Technology, under
contract with the National Aeronautics and Space Administration.

{\it Facilities:} \facility{Keck:I (LRISp)}
 


\end{document}